# Theoretical investigation of the HgF radical towards laser cooling and *e*EDM measurement

Zhenghai Yang, Jing Li, Qinning Lin, Liang Xu[#], Hailing Wang, Tao Yang[*] and Jianping Yin[*]

*State Key Laboratory of Precision Spectroscopy, East China Normal University, Shanghai 200062, P. R. China*

[#]Current address: *Key Laboratory for Laser Plasmas (Ministry of Education) and School of Physics and Astronomy, Shanghai Jiao Tong University, Shanghai 200240, P. R. China*

E-mail: [*]T. Y.: tyang@lps.ecnu.edu.cn; J. P. Y.: jpyin@phy.ecnu.edu.cn

# Abstract

In order to realize more sensitive $e$EDM measurement, it would be worthwhile to find some new laser-cooled molecules with larger internal effective electric field ($E_{\text{eff}}$), higher electric polarizability and longer lifetime of the $e$EDM measurement state. Here we explore the merits of mercuric monofluoride ($^{202}$Hg$^{19}$F, $X^2\Sigma_{1/2}$) for its potential of laser cooling and $e$EDM measurement. We theoretically investigated the electronic, rovibrational and hyperfine structures and verified the highly diagonal Franck-Condon factors (FCFs) of the main transitions by the Rydberg-Klein-Rees inversion (RKR) method and the Morse approximation. Hyperfine manifolds of the $X^2\Sigma_{1/2}$ ($v = 0$) rotational states were examined with the effective Hamiltonian approach and a feasible sideband modulation scheme was proposed. In order to enhance optical cycling, the microwave remixing method was employed to address all the necessary levels. The Zeeman effect and the hyperfine structure magnetic $g$ factors of the $X^2\Sigma_{1/2}$ ($v = 0, N = 1$) state were studied subsequently. Finally, its statistical sensitivities for the $e$EDM measurement were estimated respectively to be about $9\times10^{-31}$ $e$ cm (the laser cooled transverse beam experiment), $2\times10^{-31}$ $e$ cm (the fountain experiment) and $1\times10^{-32}$ $e$ cm (experiment with trapped cold molecules), indicating that $^{202}$Hg$^{19}$F might be another promising $e$EDM candidate when compared with the most recent ThO result of $d_e = (4.3 \pm 3.1_{\text{stat}} \pm 2.6_{\text{syst}})\times10^{-30}$ $e$ cm (Nature, 562, 355 (2018)). In addition, the possibility of direct Stark decelerating of the HgF radical was also discussed.

## I. INTRODUCTION

The measurement of the electron's electric dipole moment (*e*EDM) has been a platform for searching new physics beyond the Standard Model of the elementary physics since 1950s, as suggested by E. M. Purcell and N. F. Ramsey [1]. A non-zero *e*EDM value can be directly used to trace the origin of the CP-violation [2]. An often-quoted value of *e*EDM in the Standard Model was predicted to be below $10^{-38}$ *e* cm, which is obviously far below the current experimental sensitivities [3-6]. However, many extensions to the Standard Model predict a much larger *e*EDM value, such as the SUSY variants and generic models [7,8], some of which are under direct test by experimental efforts with the recently published upper limit of $d_e < 1.1 \times 10^{-29}$ *e* cm [6].

Mainly due to their important features of large internal effective electric field ($E_{\mathrm{eff}}$) and high electric polarizability, there have been numerous diatomic molecules and molecular ions with heavy nuclei that are examined theoretically and experimentally to probe the *e*EDM, among which some precise experimental results have been obtained with YbF [3], PbO [9], ThO [4, 6] and HfF$^+$ [5]. Besides those, diatomic molecules and molecular ions such as PbF [10], WC [11], RaF [12], HgX [13], ThF$^+$ [14] and BaF [15] are under either theoretical investigation or experimental attempts. We stress that only the YbF [16], RaF [17] and BaF [18] radicals have been either experimentally or theoretically studied for their laser cooling capabilities, showing their competitive potentials in the *e*EDM measurement with much longer coherence time.

The large $E_{\mathrm{eff}}$, one of the features for *e*EDM precision measurement, was evaluated to be about 104 GV/cm for HgF [13], larger than ThO with 84 GV/cm [19], YbF with 26 GV/cm [20] and HfF$^+$ with 24 GV/cm [21]. In this paper, we investigate further on $^{202}$Hg$^{19}$F about its laser cooling features towards an *e*EDM measurement: (i) much longer interaction time between the molecules and the external fields due to the long lifetime of the $X^2\Sigma_{1/2}$ ground state for *e*EDM measurement compared with a metastable state; (ii) the highly diagonal Franck-Condon (FC) matrix between the electronic ground state and the energetically electronically excited $C^2\Pi_{1/2}$ state; (iii) a simple hyperfine structure in $X^2\Sigma_{1/2}$ and strong spontaneous radiation decay rate ($\Gamma \approx 2\pi \times 23$ MHz) because of short lifetime of excited $C^2\Pi_{1/2}$ state ($\tau \approx 6.93$ ns) [22]. Considering the dissociative nature of the unbound $A^2\Pi^+$ state, there is only one intermediate electronic state $B^2\Sigma^+$ that participates in the allowed transitions. Here, we choose $C^2\Pi_{1/2} \leftarrow X^2\Sigma_{1/2}$ as a cooling channel for the vast difference between the decay rates $\gamma_{CX}$ and $\gamma_{CB}$ (the transition frequency $\omega_{CX}/\omega_{CB} \approx 3$ and the transition dipole moment $d_{CX}/d_{CB} \approx 20$, $\gamma \propto \omega^3 d^2$, thus $\gamma_{CB}/\gamma_{CX} \approx 10^{-4}$ [22]). Therefore the existence of the intermediate state and the unwanted $B^2\Sigma^+ \leftarrow C^2\Pi_{1/2}$ leak will not limit laser cooling process significantly, but only make the optical cycling more complex. YO [23] and BaF [18] radicals were processed in the similar way for the existence of an intermediate electronic state. The lifetime of $B^2\Sigma^+$ state is also very short ($\tau \approx 8.05$ ns), and the undesired leakage of $B^2\Sigma^+ \leftarrow C^2\Pi_{1/2}$ will decay rapidly to $|X, N = 0, 2, +\rangle$ because of selection rules, however, we can close these additional loss channels by microwave mixing of rotational states.

## II. THE VIBRATIONAL TRANSITIONS AND FRANCK-CONDON FACTORS BETWEEN $X^2\Sigma_{1/2}$ AND $C^2\Pi_{1/2}$ STATES

The vibrational branching ratios of a molecular system are represented by the FCFs among the involved optical transitions. As illustrated in FIG. 1, we depicted the laser scheme and spontaneous decay based on the calculated transition wavelength $\lambda_{vv'}$ and the corresponding FCFs $f_{v'v}$. The molecular parameters of the states $X^2\Sigma_{1/2}$ and $C^2\Pi_{1/2}$ used in the calculation are listed in Table I.

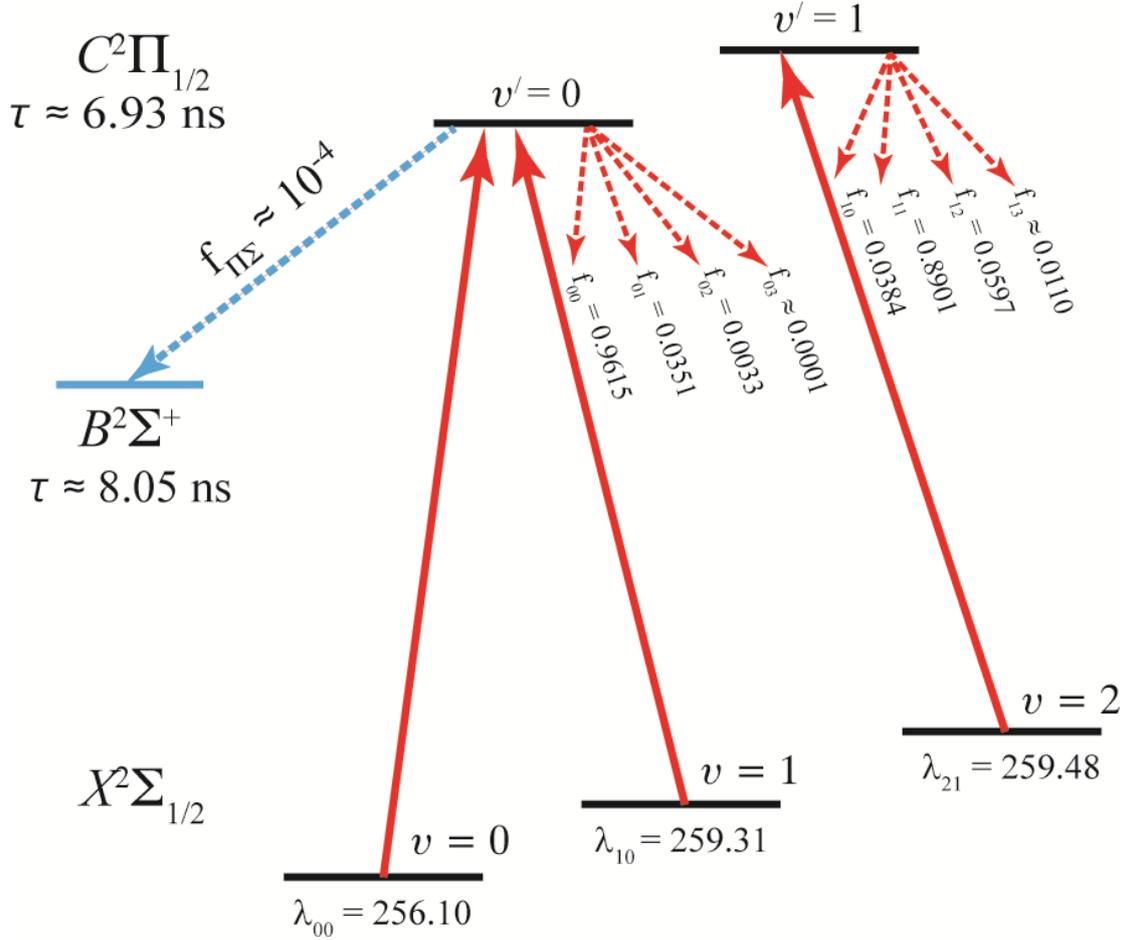

FIG. 1. The proposed scheme to create a quasicycling transition for laser cooling of HgF. Solid black lines indicate the relevant electronic and vibrational level structure. Solid upward lines indicate the laser-driven transitions at the wavelengths $\lambda_{vv'}$. Solid dotted red lines indicate the spontaneous decays from the $C$ state along with FCFs $f_{v'v}$. The dotted blue line indicates the undesired leakage of $B^2\Sigma^+ \leftarrow C^2\Pi_{1/2}$.

The FCFs of $C^2\Pi_{1/2} \leftarrow X^2\Sigma_{1/2}$ transition were calculated with RKR (Rydberg-Klein-Rees) method [24] and Morse potential [25]. For the Morse potential method, one-dimensional analytical potential function is constructed using $U(x) = D[(1 - e^{-\beta x})^2 - 1]$ where $D = \hbar\omega_e^2/4\omega_e\chi_e$ and $\omega_e\chi_e = \beta^2\hbar/2\mu$ (the potential is characterized by the depth $D$ and the range $\beta$; $x$, $\mu$, $\omega_e$ and $\omega_e\chi_e$ represent the position of the equilibrium point, the reduced mass, the standard harmonic and anharmonic spectroscopic parameters, respectively). For the RKR method, the related potential energy curves are numerically modeled by calculating classical turning points with vibrational and rotational spectroscopic constants.

TABLE I. Parameters for the involved electronic states of HgF.

| Molecular parameters | $X^2\Sigma_{1/2}$ | $C^2\Pi_{1/2}$ |
|---|---|---|
| $T_e$ (cm$^{-1}$) | 0 | 39060 [26] |
| $\omega_e$ (cm$^{-1}$) | 490.8 [26] | 468.6 [22] |
| $\omega_e\chi_e$ (cm$^{-1}$) | 4.05 [26] | 10.33 [22] |
| $r_e$ (Å) | 2.110 [22] | 2.092 [22] |
| $\tau$ (ns) | - | 6.93 [22] |

The overlap integral $\langle v|v'\rangle$ and FCFs $\langle v|v'\rangle^2$ were calculated with respect to the wave functions of each involved vibrational state. Some of the FCFs are listed in Table II and their corresponding transition wavelengths were shown in Table III. The sum of $f_{00}$, $f_{01}$, and $f_{02}$ is close to the unity (~0.9999 for either method), which is very similar to the calculated results reported in Ref. [22] that the sum of $f_{00}$, $f_{01}$, and $f_{02}$ is also close to unity (> 0.9999), so almost $10^4$ photons can be scattered to slow the molecules. In fact, if the fourth laser beam is used, nearly $10^5$ photons will be scattered (the sum of $f_{00}$, $f_{01}$, $f_{03}$ and $f_{04}$ is about 0.99999).

TABLE II. The calculated FCFs of HgF by the Morse potential method and the RKR inversion method.

| Methods | $f_{00}$ | $f_{01}$ | $f_{02}$ | $f_{11}$ |
|---|---|---|---|---|
| Morse potential | 0.9760 | 0.0204 | 0.0035 | 0.9540 |
| RKR inversion | 0.9615 | 0.0351 | 0.0033 | 0.8901 |

Based on the calculated results, $C^2\Pi_{1/2}$ ($v'= 0$) ← $X^2\Sigma_{1/2}$ ($v = 0$) transition was chosen as the main cooling transition due to its favorable FCFs ($f_{00}$ = 0.9615), while the cooling laser wavelength is $\lambda_{00}$ = 256.10 nm. The vibrational leakages can be addressed by repumping the $X^2\Sigma_{1/2}$ ($v = 1$) directly to the $C^2\Pi_{1/2}$ ($v' = 0$) as well as $v = 2$ to $v' = 1$ with wavelengths $\lambda_{10}$ = 259.31 nm and $\lambda_{21}$ = 259.48 nm respectively. Theoretical results are in good agreement with experimental ones, as listed in Table III.

TABLE III. The comparison between the calculated and experimental results of the transition wavelengths of the $X^2\Sigma_{1/2}$ and $C^2\Pi_{1/2}$ states of HgF.

| Transitions wavelength | Theoretical (nm) | Experimental (nm) |
|---|---|---|
| $\lambda_{00}$ | 256.10 | 256.06 [26] |
| $\lambda_{10}$ | 259.31 | 259.24 [26] |
| $\lambda_{21}$ | 259.48 | 259.17 [26] |

## III. HYPERFINE STRUCTURE (HFS) OF THE $^{202}$Hg$^{19}$F MOLECULE

In this section, we will discuss the HFS of the lowest rotational levels of the $X^2\Sigma_{1/2}$ state and the elimination of dark rotational states by the microwave remixing method. The discussion is very important for implementing nearly closed optical transitions. The $X^2\Sigma_{1/2}$ state of HgF is a Hund's case (b) state while $C^2\Pi_{1/2}$ is a Hund's case (a) state, therefore $N$ is a good quantum number for $X^2\Sigma_{1/2}$ but $J$ is a good quantum number for $C^2\Pi_{1/2}$. Considering the angular momentum and parity selection rules, the parities of the initial and final states of the transition should be opposite and $\Delta J = 0, \pm 1$. As a result, driving $|C, v' = 0, J' = 1/2, +\rangle \leftarrow |X, v = 0, N = 1, -\rangle$ transition will allow a spontaneous decay that will only go back to $N = 1$ state.

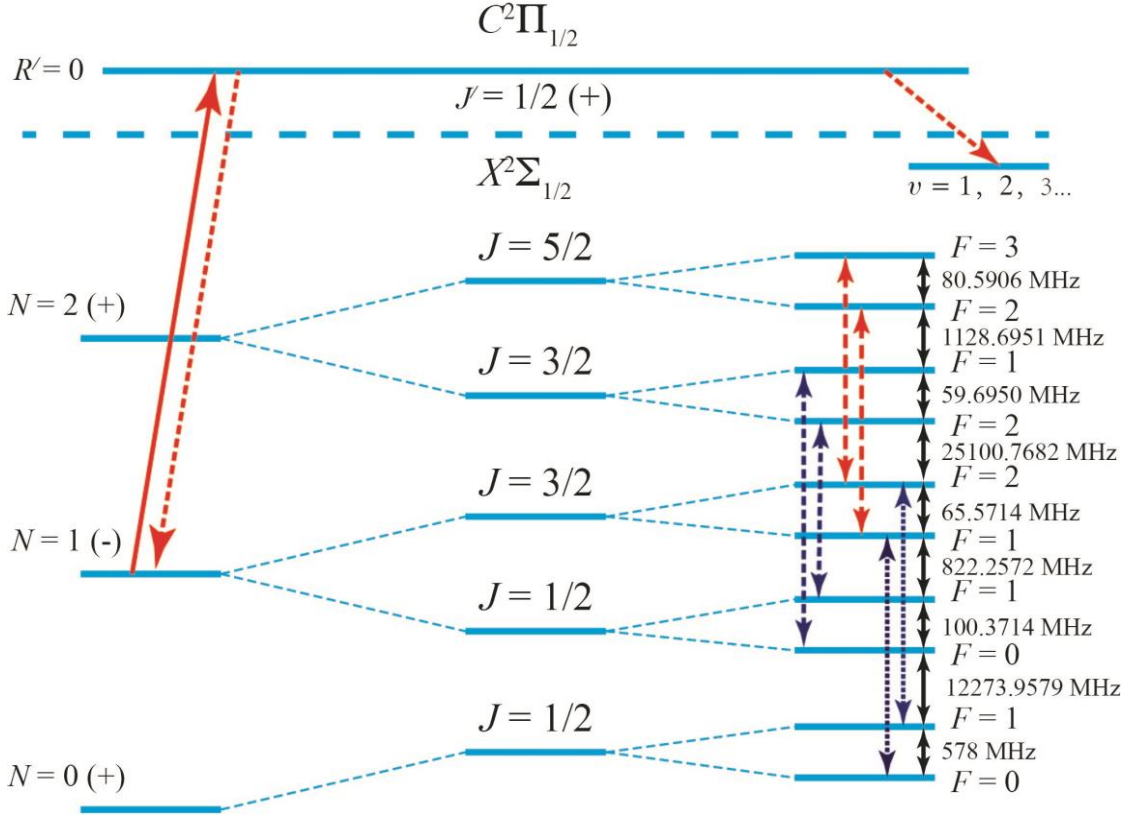

FIG. 2. Driving $|C, v' = 0, J' = 1/2, +\rangle \leftarrow |X, v = 0, N = 1, -\rangle$ transition will allow a spontaneous decay that only goes back to $N = 1$ state. Unwanted $B^2\Sigma^+ \leftarrow C^2\Pi_{1/2}$ transition will eventually end up with the $|X, N = 0,2, +\rangle$ state (not shown in the figure). The $\Delta J = +1$, $\Delta F = +1$ microwave mixing transitions is used to close additional loss channels.

For the $^{202}$Hg$^{19}$F molecule, the total angular momentum operator $\hat{F} = \hat{N} + \hat{S} + \hat{I}$ in the ground $X^2\Sigma_{1/2}$ state. Since $I_{Hg} = 0$ and $I_F = 1/2$, the spin-rotational and hyperfine interactions can split the $|X, v = 0, N = 1\rangle$ state into four sublevels. For a nearly closed transitions scheme, all of the four hyperfine levels should be pumped simultaneously in order to prevent molecules from accumulating into one state. This can be realized by sideband modulation with commercial Electro-optic modulators (EOMs).

The effective Hamiltonian describes all the involved inter-couplings degrees of freedom in a molecular system. For the $^{202}$Hg$^{19}$F $X^2\Sigma_{1/2}$ state in particular, the effective Hamiltonian ($H_{\text{eff}}$) contains the molecular rotational term $H_R$, the spin-rotational coupling $H_{SR}$, and the hyperfine interaction $H_{hfs}$. By the aid of the Frosch and Foley constants [27], the following expressions can be derived:

$$H_R = B_v \hat{N}^2 - D_v \hat{N}^4,$$

$$H_{SR} = Y_v T^1(\hat{S}) T^1(\hat{N}),$$

$$H_{hfs} = b_v T^1(\hat{I}) T^1(\hat{S}) + c_v T^1_{q=0}(\hat{I}) T^1_{q=0}(\hat{S}) + C_{vN} T^1(\hat{I}) T^1(\hat{N}),$$

$$b_{Fv} = b_v + c_v/3,$$

$$H_{eff} = H_R + H_{SR} + H_{hfs}. \tag{1}$$

Here $B_v$, $D_v$, $Y_v$, $b_{Fv}$, and $c_v$ represent the molecular rotational constant, the centrifugal distortion constant, the spin-rotational coupling constant, the Fermi contact interaction constant and the dipole-dipole interaction constant respectively. $C_{vN}$ is negligibly small compared with other constants. The hyperfine parameters $A_\perp$ and $A_\parallel$ (195 MHz and 1344 MHz) were measured by Knight *et al.* [28]. Since $b_v = A_\perp$ and $c_v = A_\parallel - A_\perp$, the values of $b_v$ and $c_v$ were derived to be 195 MHz ($6.5 \times 10^{-3}$ cm$^{-1}$) and 1149 MHz (0.0383 cm$^{-1}$) respectively. Rotational and hyperfine structure parameters as well as the electric dipole moment for $X^2\Sigma_{1/2}$ ($v = 0$) state of $^{202}$Hg$^{19}$F are listed in Table IV.

TABLE IV. Rotational and hyperfine structure parameters and the electric dipole moment for the $X^2\Sigma_{1/2}$ ($v = 0$) state of $^{202}$Hg$^{19}$F.

| Molecular parameters | $X^2\Sigma_{1/2}$ | References |
|---|---|---|
| $B_v$ (cm$^{-1}$) | 0.2181 | [22] |
| $D_v$ (cm$^{-1}$) | $3.5081 \times 10^{-7}$ | [22] |
| $Y_v$ (cm$^{-1}$) | 0.0143 | [29] |
| $b_v$ (cm$^{-1}$) | $6.5 \times 10^{-3}$ | [28] |
| $c_v$ (cm$^{-1}$) | 0.0383 | [28] |
| $\mu_e$ (Debye) | 4.15 | [30] |

With all the parameters mentioned above, the corresponding matrix elements for each term of the $H_{eff}$ with the basis $|N, S, J, I, F, m_F\rangle$ are derived as follows.

$$\langle N', S, J', I, F', m'_F | B_v \hat{N}^2 - D_v \hat{N}^4 | N, S, J, I, F, m_F \rangle = \delta_{N'N} \delta_{J'J} \delta_{F'F} \delta_{m'_F m_F} N(N+1)[B_v - D_v N(N+1)], \tag{2}$$

$$\langle N',S,J',I,F',m'_F|Y_v T^1(\hat{S})T^1(\hat{N})|N,S,J,I,F,m_F\rangle = \delta_{N'N}\delta_{J'J}\delta_{F'F}\delta_{m'_F m_F}Y_v(-1)^{N+J+S}[S(S+1)(2S+1)]^{1/2}[N(N+1)(2N+1)]^{1/2}\begin{Bmatrix} S & N & J \\ N & S & 1 \end{Bmatrix}, \quad (3)$$

$$\langle N',S,J',I,F',m'_F|b_v T^1(\hat{I})T^1(\hat{S})|N,S,J,I,F,m_F\rangle =$$

$$\delta_{N'N}\delta_{F'F}\delta_{m'_F m_F}b_v(-1)^{J'+F+I+J+N+1+S}[(2J'+1)(2J+1)]^{1/2}[S(S+1)(2S+1)]^{1/2}[I(I+1)(2I+1)]^{1/2}\begin{Bmatrix} I & J' & F \\ J & I & 1 \end{Bmatrix}\begin{Bmatrix} J & S & N \\ S & J' & 1 \end{Bmatrix}, \quad (4)$$

$$\langle N',S,J',I,F',m'_F|c_v T^1_{q=0}(\hat{I})T^1_{q=0}(\hat{S})|N,S,J,I,F,m_F\rangle =$$

$$\delta_{N'N}\delta_{F'F}\delta_{m'_F m_F}(-\sqrt{30}/3)c_v(-1)^{J'+F+I+N}[(2J'+1)(2J+1)]^{1/2}[S(S+1)(2S+1)]^{1/2}[I(I+1)(2I+1)]^{1/2}(2N+1)\begin{pmatrix} N & 2 & N \\ 0 & 0 & 0 \end{pmatrix}\begin{Bmatrix} I & J' & F \\ J & I & 1 \end{Bmatrix}\begin{Bmatrix} J & J' & 1 \\ N & N & 2 \\ S & S & 1 \end{Bmatrix}, \quad (5)$$

$$\langle N',S,J',I,F',m'_F|C_{vN}T^1(\hat{I})T^1(\hat{N})|N,S,J,I,F,m_F\rangle =$$

$$\delta_{N'N}\delta_{F'F}\delta_{m'_F m_F}C_{vN}(-1)^{2J+F'+I+N'+S+1}[N(N+1)(2N+1)]^{1/2}[I(I+1)(2I+1)]^{1/2}[(2J'+1)(2J+1)]^{1/2}\begin{Bmatrix} I & J & F' \\ J' & I & 1 \end{Bmatrix}\begin{Bmatrix} N & J & S \\ J' & N' & 1 \end{Bmatrix}. \quad (6)$$

Eigenvalues and eigenvectors were obtained by numerical diagonalization of the effective Hamiltonian matrix representations of the $X^2\Sigma_{1/2}$ state. Once the HFS of $N = 1$ state is clear, sideband modulation scheme for the pumping laser to cover all the four hyperfine levels of $N = 1$ level simultaneously can be proposed.

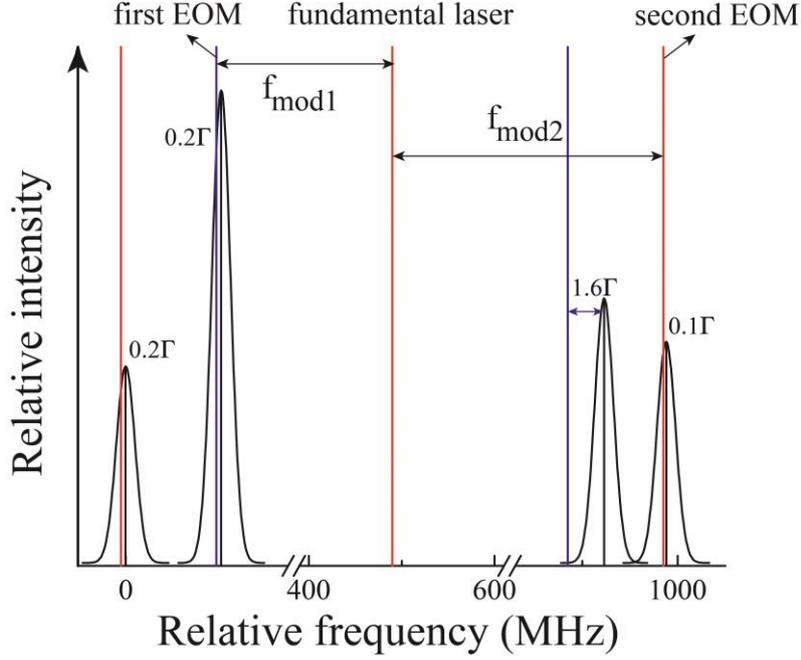

FIG. 3. The proposed sideband modulation scheme to simultaneously cover all four hyperfine levels of $|X, v = 0, N = 1\rangle$ state. The calculated molecular fluorescence spectra (black curved line) are shown with its natural linewidth and central frequencies (solid black lines). The relative intensity represents the branching ratios from the $|C, v' = 0, J' = 1/2, +\rangle$ state to each hyperfine level. Two EOMs are used in the scheme with modulation frequency of $f_{\text{mod1}}$ = 395 MHz (blue) and $f_{\text{mod2}}$ = 495 MHz (red), respectively [31].

For the $C^2\Pi_{1/2} \leftarrow X^2\Sigma_{1/2}$ transition of $^{202}$Hg$^{19}$F, decay rate $\Gamma = 2\pi \times 23$ MHz, and saturation irradiance $I_S = \pi hc\Gamma/(3\lambda^3) \approx 180$ mW/cm$^2$. As shown in FIG. 3, the four hyperfine levels of $N = 1$ are all addressed with detuning within $2\Gamma$ to the respective peaks.

As for the undesired leak, the leakage decay to the intermediate electronic state $B^2\Sigma^+$ will end up going back to $|X, v = 0, N = 0, 2, +\rangle$ due to selection rules, and the microwave remixing method will be used to eliminate this leakage, similar to Ref.s [23,32]. Microwave radiation tuned to $f_0 = \sim 13$ GHz can drive $|N = 0, F = 0\rangle \leftrightarrow |N = 1, J = 3/2, F = 1\rangle$ and $|N = 0, F = 1\rangle \leftrightarrow |N = 1, J = 3/2, F = 2\rangle$ transitions to mix the $N = 0$ and $N = 1$ hyperfine levels, while the $N = 2$ is remixed to

$N = 1$ just by doubling the frequency $f_0$ to drive $\Delta J = +1$, $\Delta F = +1$ transitions.

TABLE V. Calculated frequencies for $\Delta J$=+1, $\Delta F$=+1 hyperfine transitions in the lowest rotational levels of the $X^2\Sigma_{1/2}$ state.

| $N' - N$ | $J' - J$ | $F' - F$ | $f$ (MHz) |
|---|---|---|---|
| 1-0 | 3/2-1/2 | 1-0 | 13774.5865 |
|  |  | 2-1 | 13262.1579 |
| 2-1 | 3/2-1/2 | 1-0 | 26148.6632 |
|  |  | 2-1 | 25988.5968 |
|  | 5/2-3/2 | 2-1 | 26354.7297 |
|  |  | 3-2 | 26369.7489 |

## IV. BRANCHING RATIOS FOR THE $C^2\Pi_{1/2} \leftarrow X^2\Sigma_{1/2}$ TRANSITION

The distribution of the laser intensity can be determined by the branching ratios which reflect the transition strengths for all the hyperfine decay paths. In order to calculate the branching ratios, $J$ mixing for the electric dipole transitions in $C^2\Pi_{1/2} \leftarrow X^2\Sigma_{1/2}$ was considered first. The calculations were based on the equations (7) and (8), and the details are described in Ref. [33].

$$|F = N^\pm, M\rangle = x^\pm |J = N + \tfrac{1}{2}, F = N, M\rangle + y^\pm |J = N - \tfrac{1}{2}, F = N, M\rangle, \tag{7}$$

$$\frac{x^\pm}{y^\pm} = -\frac{\langle J=N+\tfrac{1}{2}, F=N, M | H | J=N-\tfrac{1}{2}, F=N, M\rangle}{\langle J=N+\tfrac{1}{2}, F=N, M | H | J=N+\tfrac{1}{2}, F=N, M\rangle - E_N^\pm}. \tag{8}$$

Here $x$ and $y$ represent the coefficients of the superposition of pure $J$ states. As shown in Table VI, $J$ mixing exists only in $F = 1$ of the $N = 1$ manifold, and the $g$ factors listed are only valid for magnetic field that induces small energy shift comparable to the hyperfine structure.

TABLE VI. The $g$ factors of the $X^2\Sigma_{1/2}$ ($v = 0$) state.

| Mixed lable | Superposition of pure $J$ states | $g$ (without $J$ mixing) | $g$ (with $J$ mixing) |
|---|---|---|---|
| $|J = 1/2, F = 0\rangle$ | $|J = 1/2, F = 0\rangle$ | 0.00 | 0.00 |
| $|J = 1/2, F = 1\rangle$ | $0.8575|J = 1/2, F = 1\rangle - 0.5145|J = 3/2, F = 1\rangle$ | -0.33 | -0.44 |
| $|J = 3/2, F = 1\rangle$ | $0.5145|J = 1/2, F = 1\rangle + 0.8575|J = 3/2, F = 1\rangle$ | 0.83 | 0.94 |
| $|J = 3/2, F = 2\rangle$ | $|J = 3/2, F = 2\rangle$ | 0.50 | 0.50 |

The calculation of the electric dipole transitions of $C^2\Pi_{1/2} \leftarrow X^2\Sigma_{1/2}$ is based on Hund's case (a) basis $|\Lambda, S, \Sigma, \Omega, J, I, F, m_F\rangle$. According to Equation (9), the pure $J$ states of Hund's case (b) $X$ state can be converted to Hund's case (a) basis, and Hund's case (a) $A$ state can be expressed by Equation (10). The electric dipole matrix elements were then calculated by Equation (11) and $T^{(1)}(d)$ is the electric-dipole operator written in the spherical tensor. More details are described in Ref. [34] (Equation 6.149 and 6.234). Similar method was also used for BaF [32] and MgF [35].

$$|\Lambda, N, S, J\rangle = \sum_\Omega \sum_\Sigma (-1)^{J+\Omega} \sqrt{2N+1} \begin{pmatrix} S & N & F \\ \Sigma & \Lambda & -\Omega \end{pmatrix} |\Lambda, S, \Sigma, \Omega, J, F\rangle, \tag{9}$$

$$|\Lambda^s, J, M, \pm\rangle = \frac{1}{\sqrt{2}}(|\Lambda^s, S, \Sigma, J, \Omega, M\rangle \pm (-1)^{J-S}|-\Lambda^s, S, -\Sigma, J, -\Omega, M\rangle), \tag{10}$$

$$m_{ij} = \langle i|T^{(1)}(d)|j\rangle = \sum_{p=-1}^{1}(-1)^{F'-m'_F}\begin{pmatrix} F' & 1 & F \\ -m'_F & p & m_F \end{pmatrix}(-1)^{F+J'+I+1}\sqrt{(2F'+1)(2F+1)}\begin{Bmatrix} J & F & I \\ F' & J' & 1 \end{Bmatrix} \times$$

$$\sum_{q=-1}^{1}(-1)^{J'-\Omega'}\begin{pmatrix} J' & 1 & J \\ -\Omega' & q & \Omega \end{pmatrix}\sqrt{(2J'+1)(2J+1)} \times \langle\Lambda', S, \Sigma'|T_q^{(1)}(d)|\Lambda, S, \Sigma\rangle. \tag{11}$$

The branching ratios for decays from hyperfine sublevels in $|A, J = 1/2, +\rangle$ to hyperfine sublevels in $|X, N = 1, -\rangle$ are listed in Table VII.

TABLE VII. Calculated hyperfine branching ratios for decays from $C^2\Pi_{1/2}$ ($J = 1/2$, +) to $X^2\Sigma_{1/2}$ ($N = 1$,-) state of HgF.

| | | | $F' = 0$ | $F' = 1$ | | |
|---|---|---|---|---|---|---|
| $J$ | $F$ | $m_F$ | $m'_F = 0$ | $m'_F = -1$ | $m'_F = 0$ | $m'_F = 1$ |
| 3/2 | 2 | -2 | 0.0000 | 0.1667 | 0.0000 | 0.0000 |
| | | -1 | 0.0000 | 0.0833 | 0.0833 | 0.0000 |
| | | 0 | 0.0000 | 0.0278 | 0.1111 | 0.0278 |
| | | 1 | 0.0000 | 0.0000 | 0.0833 | 0.0833 |
| | | 2 | 0.0000 | 0.0000 | 0.0000 | 0.1667 |
| 3/2 | 1 | -1 | 0.0019 | 0.1486 | 0.1486 | 0.0000 |
| | | 0 | 0.0019 | 0.1486 | 0.0000 | 0.1486 |
| | | 1 | 0.0019 | 0.0000 | 0.1486 | 0.1486 |
| 1/2 | 1 | -1 | 0.3315 | 0.1014 | 0.1014 | 0.0000 |
| | | 0 | 0.3315 | 0.1014 | 0.0000 | 0.1014 |
| | | 1 | 0.3315 | 0.0000 | 0.1014 | 0.1014 |
| 1/2 | 0 | 0 | 0.0000 | 0.2222 | 0.2222 | 0.2222 |

# V. THE INTERACTION OF THE EXTERNAL MAGNETIC FIELD WITH HYPERFINE LEVELS OF HgF

In order to study the features of HgF MOT, it is crucial to analyze the effects of the external magnetic field upon the HgF $X^2\Sigma_{1/2}$ hyperfine levels. The Hamiltonian and matrix expression of the Zeeman interaction are given by Equations (12) and (13).

$$\hat{H}_{\text{Zeeman}} = g_s\mu_B T^1(\hat{S})T^1_{p=0}(\hat{B}) + g_L\mu_L T^1(\hat{L})T^1_{p=0}(\hat{B}) - g_I\mu_N T^1(\hat{I})T^1_{p=0}(\hat{B}), \quad (12)$$

$$\langle N',S,J',I,F',m'_F | g_s\mu_B T^1(\hat{S})T^1_{p=0}(\hat{B}) | N,S,J,I,F,m_F \rangle =$$

$$\delta_{N'N}\delta_{m'_F m_F} g_s\mu_B B_Z (-1)^{F-m_F+F'+2J+I+N+S}[(2J+1)(2J'+1)(2F+1)(2F'+1)]^{1/2}[S(S+1)(2S+1)]^{1/2} \times \begin{Bmatrix} J & S & N \\ S & J' & 1 \end{Bmatrix} \begin{Bmatrix} F & J & I \\ J' & F' & 1 \end{Bmatrix} \begin{pmatrix} F & 1 & F' \\ -m_F & 0 & m_F \end{pmatrix}. \quad (13)$$

For the Zeeman term of the $X^2\Sigma_{1/2}$ state, $\Lambda = 0$, $\mu_B$, and $\mu_N$ represent the Bohr magneton and nuclear magneton with $\mu_B/\mu_N = 1836$, while $g_S$, $g_L$ and $g_I$ are the electron, electron orbital and nuclear $g$ factors with the values of 2.002, 1 and 5.585 respectively. To sum up, only the first term of Equation (12) is significant. Thus, the matrix representation of Zeeman effect is expressed by Equation (13), we present the HFS Zeeman shift of $X^2\Sigma_{1/2}$ ($N$ = 1) state in FIG. 4.

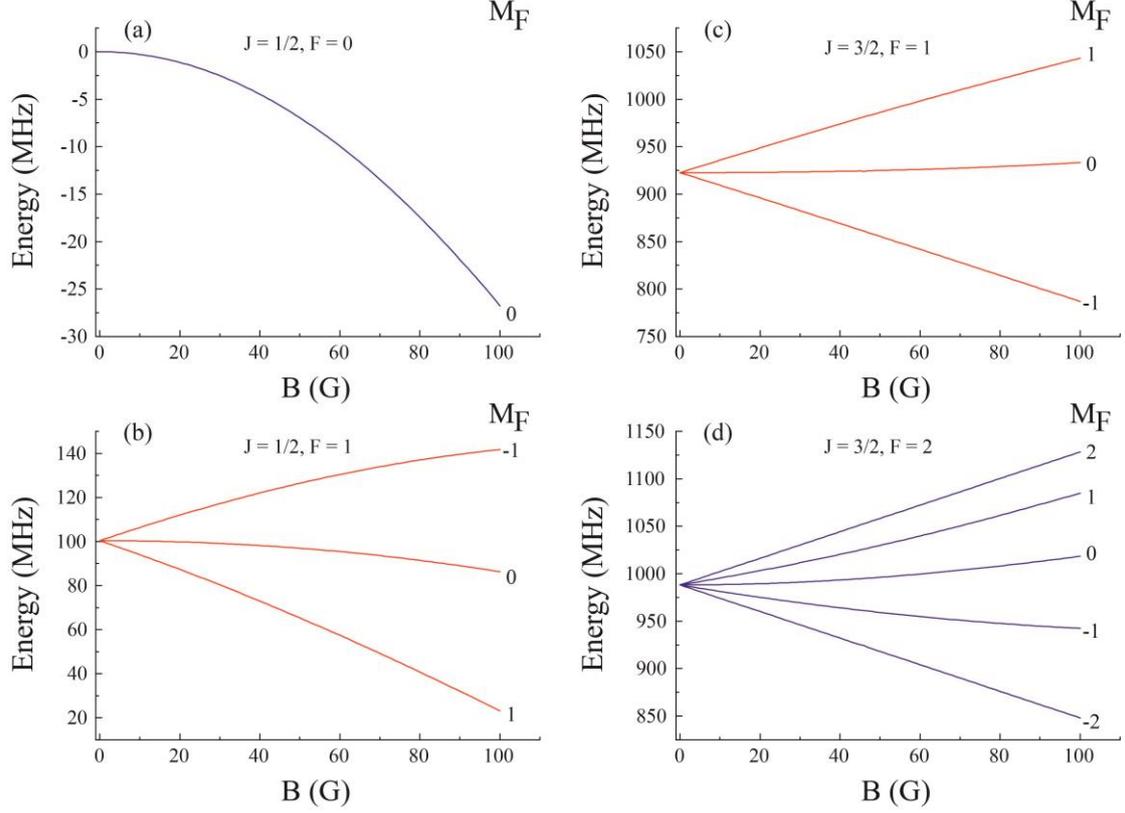

FIG. 4. The Zeeman shift of $X^2\Sigma_{1/2}$ ($N=1$): (a) for $|J=1/2, F=0\rangle$, (b) for $|J=1/2, F=1\rangle$, (c) for $|J=3/2, F=1\rangle$, and (d) for $|J=3/2, F=2\rangle$.

As shown in FIG. 4, each magnetic sublevel of the same *F* will shift completely. The *g* factor of each hyperfine structure was calculated by applying a rather small magnetic field and we adopted $g_F = \Delta U/(M_F \mu_B B)$, where *B* is the small magnetic field and $\Delta U$ is the corresponding energy difference. For the MOT experiment, the typical magnetic field is about several Gauss. For HgF, $|J=3/2, F=2\rangle$ and $|J=3/2, F=1\rangle$ states have positive *g* factors: $g_2 = 0.5$ and $g_1^+ = 0.94$, with "+" indicating the higher *J* state of the same *F*, and these states split symmetrically into eight magnetic sublevels. But the *g* factor of $|J=1/2, F=1\rangle$ turned out to be -0.44 and the *g* factor of $|J=1/2, F=0\rangle$ is even close to zero. The hyperfine structure of the HgF $X^2\Sigma_{1/2}$ ($N=1$) state used in the MOT experiment satisfies the requirement of type-II MOT system where $F' \leq F$ [36, 37].

## VI. SUITABILITY FOR *e*EDM PRECISION MEASUREMENT EXPERIMENT

For the *e*EDM measurement, the achievable statistical uncertainty can be expressed as $d_e = \hbar/(2E_{\text{eff}}\tau\sqrt{\dot{N}T})$, where $\dot{N}$ is the detected rate of the molecules, $T$ is the total integration time, $\tau$ is the interaction time of the molecules with external fields in the Ramsey interferometer. $E_{\text{eff}}$ is the internal effective electric field of the HgF $X^2\Sigma_{1/2}$ state, which is closely related to the applied electric field $E_{\text{app}}$ and its relevant polarization factor $\eta$. This polarization factor was calculated from dividing the expectation value of Hamiltonian $H_d$ by $E_{\text{app}}$. As shown in FIG. 5, if the applied electric field $E_{\text{app}}$ is 10 kV/cm, the corresponding $E_{\text{eff}}$ is 62 GV/cm, compared with YbF of 14.5 GV/cm and BaF of 9 GV/cm respectively under the same $E_{\text{app}}$.

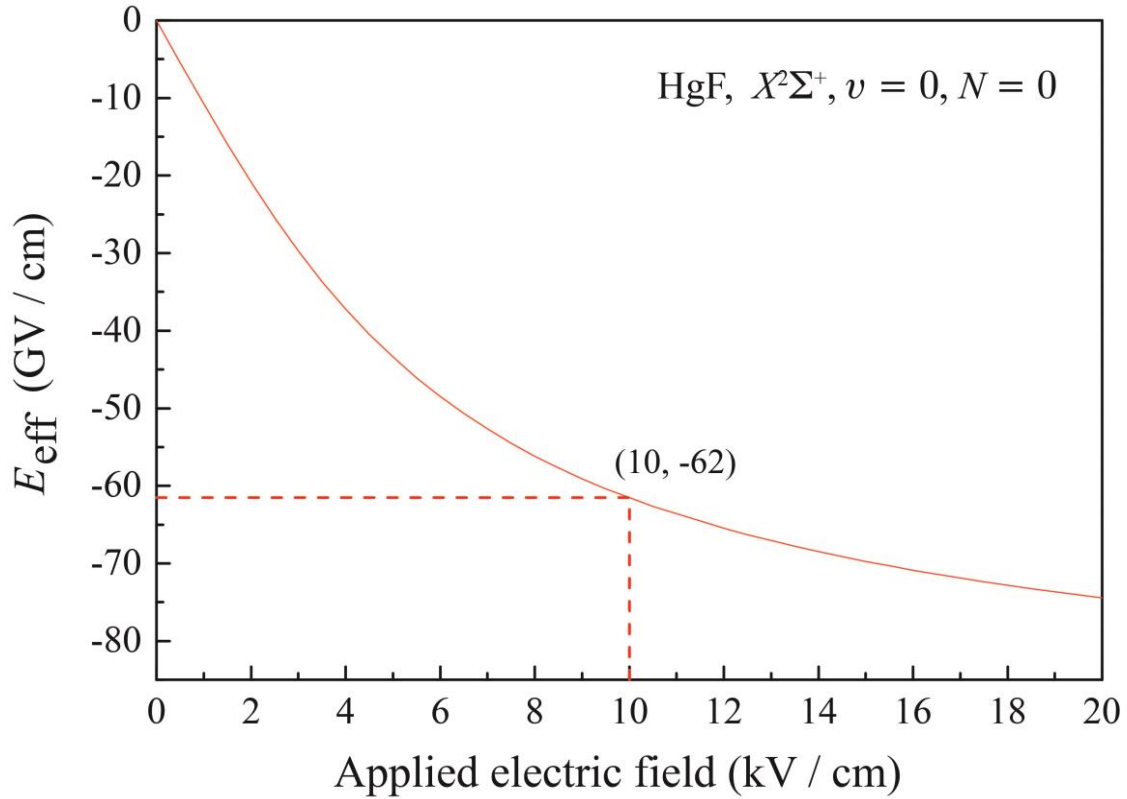

FIG. 5. Variation of $E_{\text{eff}}$ with respect to $E_{\text{app}}$ for the HgF $X^2\Sigma_{1/2}$ state. If the normal operating electric field such as $E_{\text{app}}$ = 10 kV/cm is chosen, the effective field $E_{\text{eff}}$ turns out to be about 62 GV/cm.

In order to suppress the statistical uncertainty, the increase of $\dot{N}$ can be realized by improving the molecular flux of the beam source, remixing the molecules of the other rotational states to the desired probing state before the interferometry measurement. The decrease of the forward velocity of the molecular species can effectively increase the interaction time $\tau$. For the $^{202}$Hg$^{19}$F radical, provided that a single photon at 256 nm results in a recoil velocity of 7 mm/s, and the molecular beam exiting from a two-stage buffer gas cooling source is depicted with a forward velocity of about 75 m/s [38], our calculations show that only ~ $5 \times 10^3$ photons should be scattered to slow the molecules down to ~ 40 m/s. As stated above, almost $10^4$ photons can be scattered to slow the molecules if three laser beams are used, and in principle, up to ~$10^5$ photons can be scattered if the fourth laser beam is added. However, for the horizontal beam machine, coherence interaction time is greatly limited by the gravity, and should not be longer than ~10 ms. Therefore, for the current experiment, the molecular beam with 40 m/s forward velocity is slow enough. Considering those above as well as the other parameters ever published, we are therefore able to estimate the number of molecules that is practical in the $e$EDM measurement to be ~$5 \times 10^5$ molecules per shot.

TABLE VIII. The estimated overall flux of the HgF ground state molecules that are applicable in the $e$EDM measurement with a horizontal beam machine.

| Item | Coefficient | Resulting (mol./pulse) |
|---|---|---|
| Beam flux | $5 \times 10^8$ | |
| fraction in $v = 0, N = 1$ | 0.6 (with microwave mixing) | $3 \times 10^8$ |
| Fraction of laser cooling | 0.008 | $2.4 \times 10^6$ |
| Fraction of transmission | 0.2 | $5 \times 10^5$ |

The statistical sensitivity of $^{202}$Hg$^{19}$F was then derived. The $E_{\text{eff}}$ is about 62 GV/cm with 10 kV/cm applied electric field in the laboratory, $\tau$ is about 10 ms with 40 cm long parallel plane electrodes, and $\dot{N}T$ is about $4.3 \times 10^{11}$ after one day collection of data. With all the parameters, the statistical sensitivity $d_e$ is calculated to be ~ $9 \times 10^{-31}$ $e$ cm/day$^{1/2}$.

With the developments of MOT and optical molasses methods for molecules (since then, up to ~$1\times10^5$ molecules have been loaded in the MOT [39]), new experiment schemes like the fountain experiment will show huge advantages by greatly increasing the coherence interaction time (molecules can process coherently nearly 1 s in the fountain [40], much longer than the current beam experiment limit of ~ 1 ms). In order to further suppress the statistical uncertainty of a fountain experiment, an exceptionally slow molecular beam source is needed. The upgraded two stage buffer gas cooling source will produce very intense, cold and slow molecular beams with forward velocities low enough to be brought to rest in an optical molasses or MOT, and additional longitudinal slowing is not required. For a fountain experiment, $\tau$ is about 250 ms in free fall. For a reasonable estimation, we adopted that the beam flux is estimated to be $5\times10^9$ molecules per pulse, the fraction in $v = 0$, $N = 1$ state is about 0.6 by use of microwave mixing method, the fraction of the guide is 0.065, fractions of molecules cooled by MOT and returning from the fountain to be 0.008 and 0.08 respectively, and the final detection is calculated to be ~$1.25\times10^5$ molecules per shot. Then the statistical sensitivity $d_e$ is calculated to be ~$2\times10^{-31}$ $e$ cm/day$^{1/2}$. With longer interaction time (~ 1 s) and more efficient cooling scheme, statistical sensitivity can be further suppressed to $10^{-32}$ $e$ cm/day$^{1/2}$ level. Ultimately, the most sensitive experiments will be conducted with trapped cold molecules, because up to tens of seconds interaction time may be obtained in the trapped experiments, thus the statistical sensitivity can be suppressed at least one order of magnitude. In order to achieve sensitivity at this statistical limit, the noise due to random fluctuations of the magnetic field must be suppressed to fT Hz$^{-1/2}$ level [40]. This can be achieved by good magnetic shielding together with the use of appropriate materials inside the apparatus [40].

Except for the laser cooling scheme, Stark decelerator [41,42] is also eligible, with the Stark shift of the three lowest rotational levels of the $X^2\Sigma_{1/2}$ ($v = 0$) state illustrated in FIG. 6.

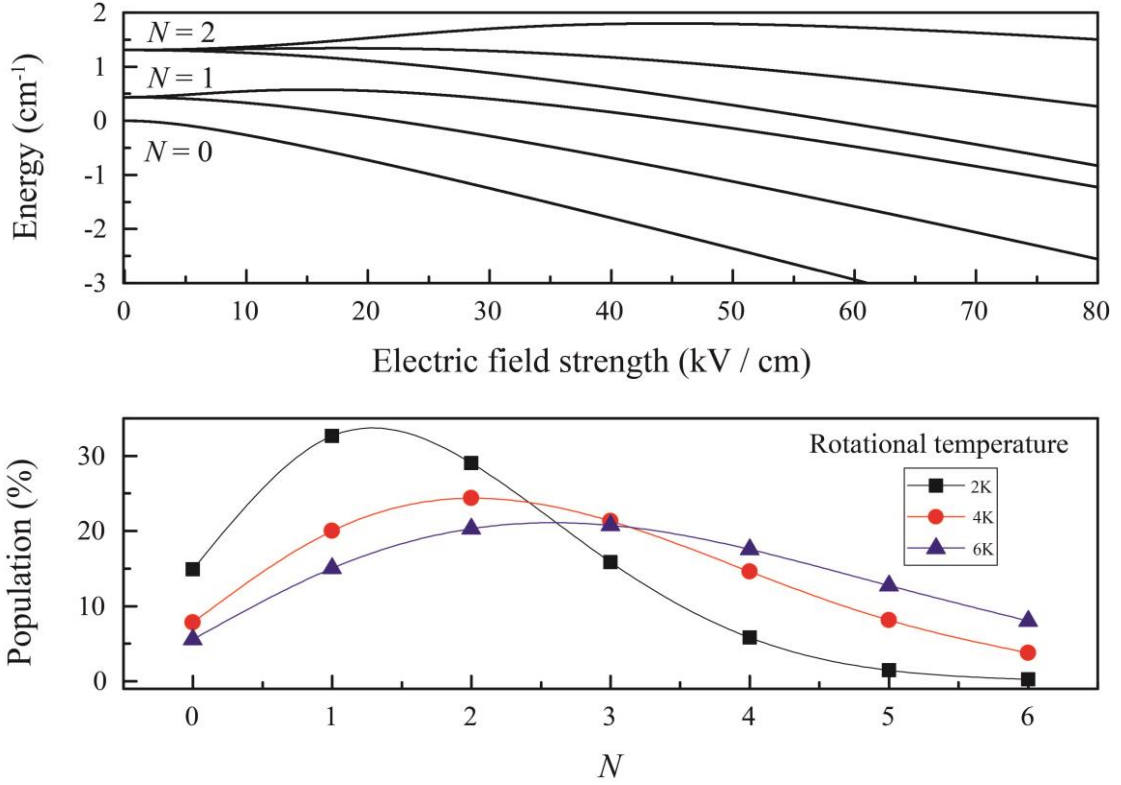

FIG. 6. The Stark shift and population of the lowest rotational levels of $X^2\Sigma_{1/2}$ ($v = 0$) state.

As shown in FIG. 6, the deceleration is the most efficient in the $N = 2$ rotational level. It is applicable that the forward velocity of the molecular beam can be decelerated to ~4 m/s before entering the interaction region by the use of the efficient Stark decelerator, and the transverse laser cooling method will be used to reduce the transverse velocity spread of the beam. The corresponding statistical uncertainty was also calculated to be at $10^{-31}$ $e$ cm/day$^{1/2}$ level.

## VII. CONCLUSION

In this paper, we have theoretically investigated the electronic, rovibrational and hyperfine structures of $^{202}$Hg$^{19}$F, and verified the highly diagonal Franck-Condon factors (FCFs) of the main transitions by the RKR method and Morse approximation. We also studied the HFS, Zeeman shift and hyperfine structure magnetic $g$ factors of its $X^2\Sigma_{1/2}$ ($N = 1$) state with the effective Hamiltonian approach. Our study indicates that the HgF radical is a good candidate for laser cooling, with less than $5\times10^3$ photons to be scattered, and the HgF radical can be longitudinally slowed from 75 m/s to 40 m/s. The statistical sensitivities of the $e$EDM measurement were estimated respectively to be about $9\times10^{-31}$ $e$ cm (the laser cooled transverse beam experiment), $2\times10^{-31}$ $e$ cm (the fountain experiment) and $1\times10^{-32}$ $e$ cm (experiment with trapped cold molecules), indicating that $^{202}$Hg$^{19}$F might be a promising $e$EDM candidate when compared with the most recent ThO result of $d_e = (4.3 \pm 3.1_{stat} \pm 2.6_{syst})\times10^{-30}$ $e$ cm. Besides, the Stark shift of the three lowest rotational levels of $X^2\Sigma_{1/2}$ ($v = 0$) state showed that HgF is also suitable for Stark deceleration, and the deceleration is most efficient in the $N = 2$ state. With an efficient Stark decelerator, the forward velocity of the molecular beam can be decelerated to ~4 m/s and the corresponding statistical sensitivity was estimated to be at the $10^{-31}$ $e$ cm/day$^{1/2}$ level.

## ACKNOWLEDGMENTS

This work is supported by the National Natural Science Foundation of China (Grants No. 91536218, 11834003, 11874151 and 11674096), the Fundamental Research Funds for the Central Universities，Shanghai Pujiang Talents Plan (18PJ1403100), Exploration funds for the Shanghai Natural Science Foundation (18ZR1412700).

Jungmann, H. L. Bethlem, and S. Hoekstra, J. Mol. Spectrosc. **300**, 22 (2014).